\documentclass[aps,pra,twocolumn,groupaddress]{revtex4-1}
\usepackage{graphicx}
\usepackage{amsmath,amssymb}
\usepackage{bm}
\usepackage{tikz}
\usepackage{textcomp}
\usetikzlibrary{positioning}

\newcommand{\ee}[1]{\relax\ifmmode 10^{#1} \else 10$^{#1}$\fi}
\newcommand{\qvalue}[1]{\left\langle #1 \right\rangle}
\newcommand{\expec}[3]{\left\langle #1 | #2 | #3 \right\rangle}
\renewcommand{\eqref}[1]{Eq.~(\ref{#1})}
\newcommand{\fgref}[1]{Fig.~\ref{fg:#1}}
\newcommand{\Fgref}[1]{Figure~\ref{fg:#1}}

\newcommand{\scref}[1]{Sec.~\ref{sc:#1}}
\newcommand{\ha}{\hat{a}}
\newcommand{\had}{\hat{a}{}^\dagger}
\newcommand{\AD}{A_\mathrm{D}}

\newcommand{\omgD}{\omega_\mathrm{D}}

\newcommand{\dd}{{\rm d}}

\def\infrac#1/#2{\leavevmode
   \textrm{\kern.1em \raise .5ex \hbox{\the\scriptfont0 #1}%
   \kern-.1em $/$%
   \kern-.1em \lower .25ex \hbox{\the\scriptfont0 #2}}}

\begin{document}
\title{Spontaneous symmetry breaking in a driven-dissipative system}
\author{J. Smits${}^1$}
\author{H.T.C. Stoof${}^2$}
\author{P. van der Straten${}^1$}
\email[]{p.vanderstraten@uu.nl}

\affiliation{$^1$Debye Institute for Nanomaterials and Center for Extreme Matter and Emergent Phenomena, Utrecht University, PO Box 80.000, 3508 TA Utrecht,The Netherlands\\
$^2$Institute for Theoretical Physics and Center for Extreme Matter and Emergent Phenomena, Utrecht University, PO Box 80.000, 3508 TA Utrecht,The Netherlands}
\date{\today}%


\pacs{03.75.Kk, 05.30.Jp, 42.25.Hz}
\maketitle


Spontaneous symmetry breaking (SSB) is a key concept in physics that for decades has played a crucial role in the description of many physical phenomena in a large number of different areas. In particle physics, for example,  the spontaneous symmetry breaking of non-Abelian symmetries provides through the Higgs mechanism the mass of $W$ and $Z$ particles, and introduces the masses of quarks and charged leptons~\cite{ryder_1996,zee_2003,kapusta_gale_2006}. 
In cosmology, SSB plays a important role in our universe through the different stages of its development, not only during the electro-weak transition just mentioned, but also during inflation~\cite{kisak_2106}. In condensed-matter physics spatial isotropy is broken spontaneously below the Curie temperature to provide a well-defined direction of magnetization to a magnetic material and the phase invariance of the macroscopic wavefunction in superfluid helium is broken below the condensation temperature, to name just two examples~\cite{negele_orland_1998,chaikin_lubensky_1995,boek_stoof}. SSB is thus an ubiquitous concept connecting several, both ``high'' and ``low'' energy, areas of physics and many textbooks describe its basic features in great detail. However, to study the dynamics of symmetry breaking in the laboratory is extremely difficult. In areas like particle physics and cosmology,  the state of matter cannot be studied by changing the control parameter and the symmetry breaking has played its role. In condensed-matter physics tiny external disturbances cause a preference for the breaking of the symmetry in a particular configuration, like a small magnetic field in ferromagnetism, and typically those disturbances cannot be avoided in experiments. Although the latter is not necessary true for a superfluid, the detection of the phase of a superfluid relies on the interferometric observation with another superfluid possessing a well-defined phase, and requires unprecedented phase-stability over long periods. Notwithstanding these complications, here we describe an experiment, in which we directly observe the spontaneous breaking of the temporal phase of a driven system with respect to the drive into two distinct values differing by $\pi$. 

\begin{figure}
\includegraphics[width=0.49\textwidth]{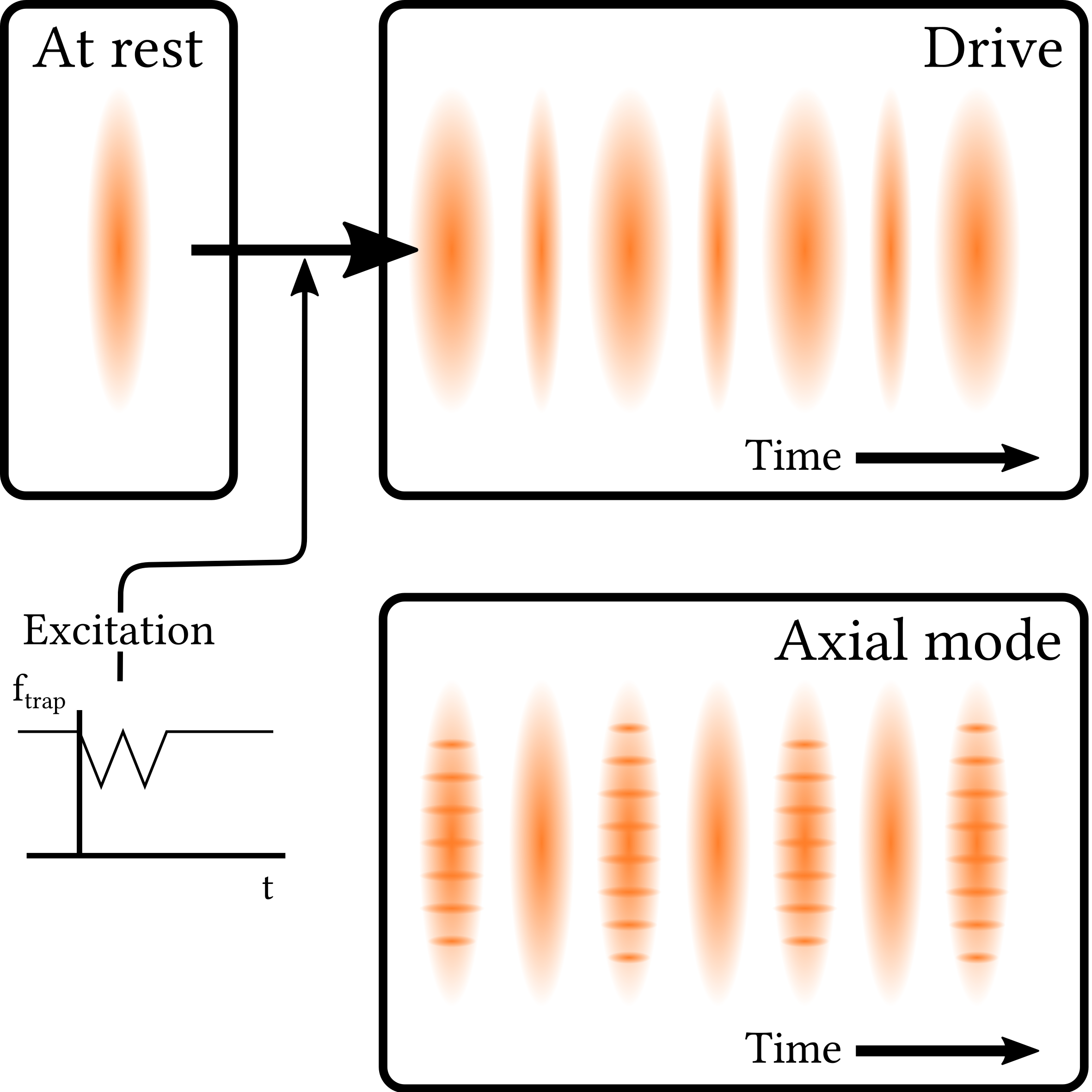}
\caption{Schematic diagram of the experiment. A superfluid droplet at rest is excited by quickly modulating the trap frequency. The superfluid starts oscillating in a breathing mode in the short (radial) direction, which acts as the drive. The drive excites in the long (axial) direction a high-order mode visible as a stripped density modulation. The oscillation of the axial mode has a phase delay with respect to the drive, which is determined up to a phase difference of $\pi$. This indeterminacy is spontaneously broken in the experiment. \label{fg:zero}}
\end{figure}

Consider a domino stone standing on its edge. In a classical world, in the absence of any external perturbation the stone will remain on its edge forever and the situation remains symmetric. In practice, it will eventually always be a small perturbation that pushes the stone either to the left or to the right. The choice for one of two directions is thus not spontaneous in this case. In a quantum world, however, the quantum domino will remain in a linear superposition falling simultaneously both to the left and right and it is not until the domino will be detected that the domino will be found on one of its two sides with exactly equal probability. The symmetry is broken and since there is no force acting on the domino stone to push the stone to a particular side, the symmetry is broken spontaneously. In our experiment we drive a superfluid droplet shaped like a cigar with an oscillatory field, which excites a high-order mode in the long direction of the droplet (see~\fgref{zero}). In the experiment a phase lag $\phi$ appears between the mode and the drive determined by the driving conditions. However, a phase lag of $\phi$ and $\phi+\pi$ are energetically identical, as the energy depends on the square of the amplitude of the mode. Whether a phase lag $\phi$ or $\phi+\pi$ is preferred, is thus undetermined at the start of the experiment. These two phases are analogues to the two sides of the domino stone. Since we can detect the shape of the superfluid non-destructively, we can detect all the collective modes of the superfluid as a function of time and thus extract the phase lag $\phi$ for any particular realization of the experiment.

The experiment is conducted in the following way. Cold atoms are trapped in a magnetic trap and evaporatively cooled to temperatures below the critical temperature for Bose-Einstein condensation~\cite{Anderson198,PhysRevLett.77.416}. The resulting Bose-Einstein condensate is a superfluid and at the temperatures in the experiment, approximately 90\% of atoms are condensed leaving 10\% as thermal atoms. The thermal atoms induce a small amount of dissipation for excitations in the superfluid. The magnetic trap is harmonic, and the trap frequency in the radial direction is much larger than the trap frequency in the axial direction. The resulting cloud of atoms will thus have an elongated, cigar-like shape. At $t=0$ the superfluid is excited by modulating the radial trap frequency by quickly ramping the current through the magnetic coils responsible for the radial confinement. This excitation induces a long-lasting oscillation of the radial size (width) of the cloud. The oscillation of the width, which is commonly referred to as the radial breathing mode~\cite{PhysRevLett.77.2360}, functions as the drive in the experiment. 

Through the non-linearity of the interactions in the superfluid different collective modes of the superfluid become coupled and in particular the drive couples to an axial mode. The drive is weak and as a result the amplitude of the axial mode is small as well. Because the coupling between  the two modes is small and the axial mode grows exponentially in time, it is only possible to detect its magnitude after a certain waiting time. After this waiting time we acquire a sequence of nearly non-destructive images of the density of the superfluid using an holographic imaging technique~\cite{smits20}. From the dynamics of the width of the superfluid we can deduce the frequency and phase of the drive with high accuracy. The axial mode is visible as an oscillation of the density profile in the axial direction and from its magnitude in time the frequency and phase of the axial mode are determined. The experiment is repeated approximately hundred times under identical initial conditions to deduce the statistics of the phase lag $\phi$ between axial mode and drive.

\begin{figure}
\includegraphics[width=0.49\textwidth]{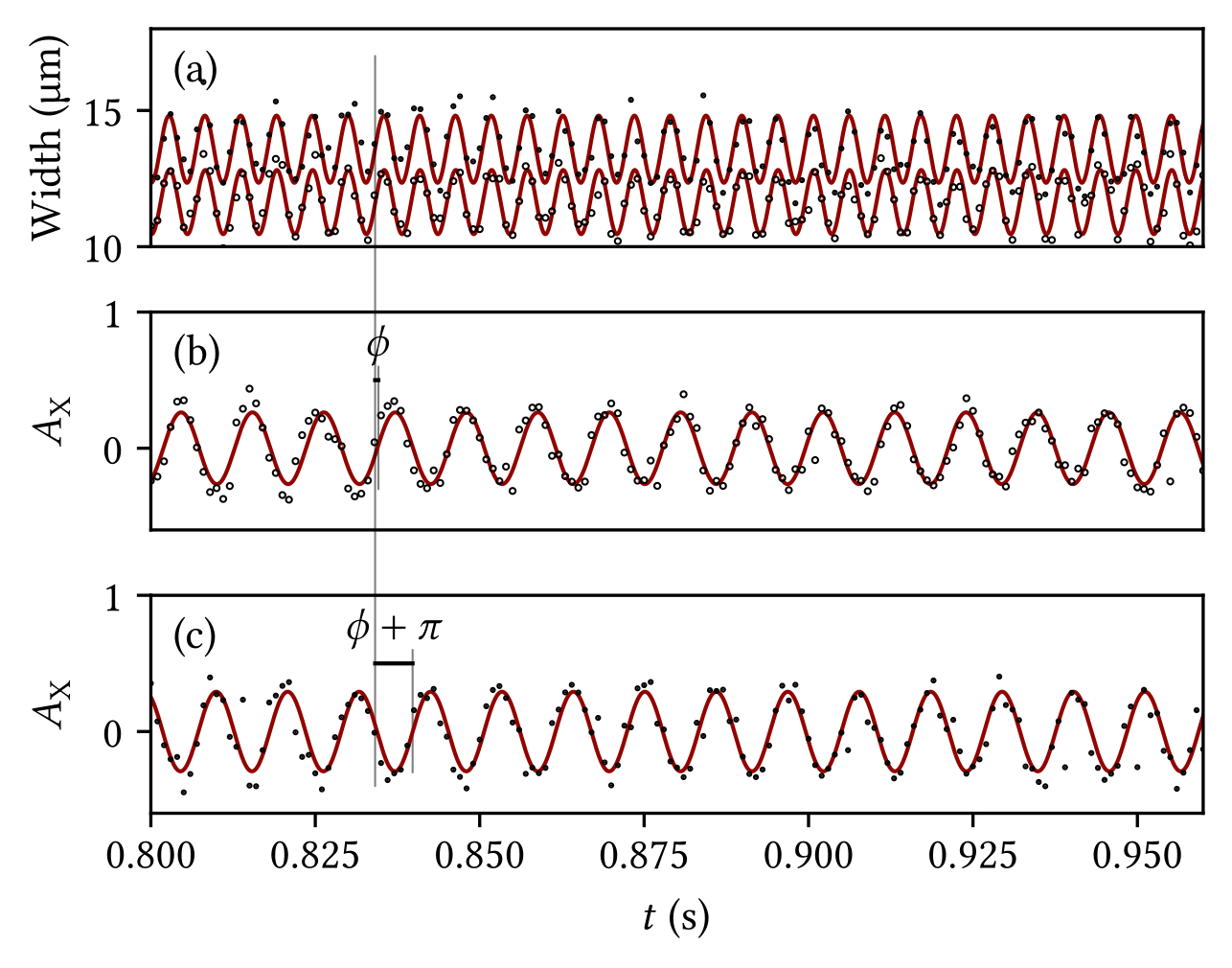}
\caption{Excitation of the modes in the experiment. {\em (a)} Radial width of the superfluid as a function of time after the kick. The two data-sets (solid and open dots) are from different sequences, where the sample has been prepared using identical parameters. For clarity, the open dots have been shifted down by 1 $\mu$m. The two curves are adjustments to the data using an oscillatory function and shows the good reproducibility of the drive. {\em (b)} and {\em (c)} Amplitude of the axial mode (dots) as a function of time after the kick, where the phase lag with the drive differs between the data in {\em (b)} and {\em (c)} by $\pi$. The solid lines are adjustments using an oscillatory function, which clearly shows the period doubling associated with the presence of a discrete time crystal~\cite{10.1146,sacha-review}. 
\label{fg:one}}
\end{figure}

Figure~\ref{fg:one} shows the results of two measurements with identical excitation of the drive. In one case the phase lag is $\phi$, whereas in the other measurement the phase lag is $\phi+\pi$. The amplitude of the axial mode is determined by fitting the mode profile of the axial mode in each image~\cite{Smits_2020}. Due to the broken discrete time symmetry in our system, the axial mode oscillates with a subharmonic of the drive~\cite{10.1146,sacha-review,Smits2018,liao19,Smits_2020}. From the data we can extract the phase lag $\phi$, as indicated by the shift of the maxima of the two oscillatory patterns. The experiment is repeated using the same parameters and a nearly identical drive is observed owing to the remarkable coherence properties of the superfluid. The axial mode, shown in \fgref{one}b,c also occurs in a reproducible manner, except for a phase lag difference  of $\pi$ between the two cases. The $\pi$ phase difference between different realizations is a result of the broken discrete time symmetry that mathematically is identical to a SSB of the $\mathbb{Z}_2$ (Ising-like) symmetry of the Hamiltonian~\cite{Smits_2020}. 

\begin{figure}
\includegraphics[width=0.49\textwidth]{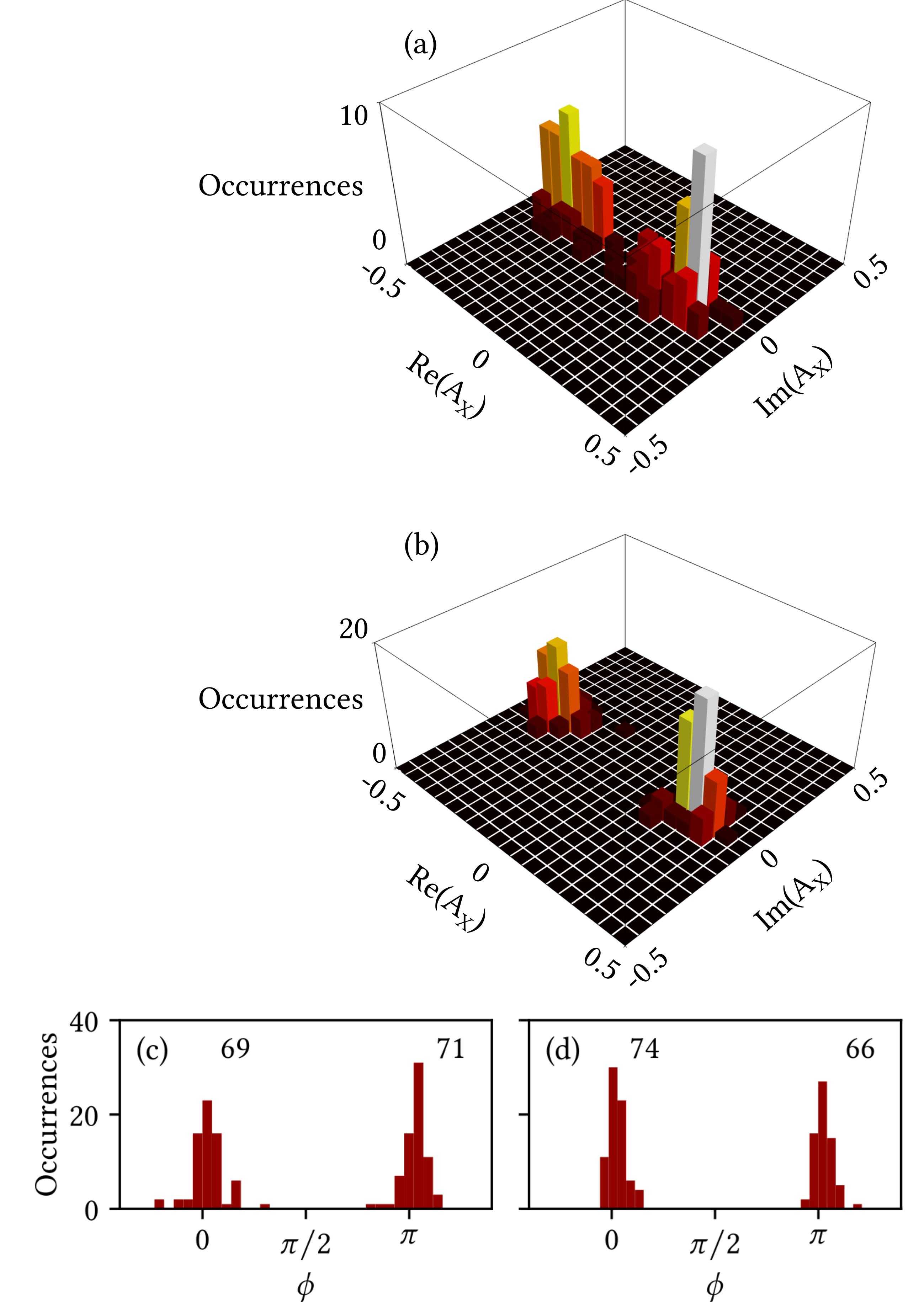}
\caption{The amplitude of the axial mode. {\em (a)} Experimental results for 140 runs, where the amplitude is represented in the complex plane with the absolute value proportional to the amplitude of the axial mode and the phase determined by the phase lag $\phi$ with the drive. {\em (b)} Result of the simulation for 140 runs with $\delta = (2.00 \pm 0.75)$ Hz, and $g = (2 \pm 1) \times $ \ee{-4} Hz. {\em (c)} and (d) Histogram of the phase lag $\phi$ of the axial mode with the drive for the experiments and simulation, respectively.  
\label{fg:two}}
\end{figure}

In \fgref{two} we plot the complex amplitude of the axial mode for different experimental runs using the same parameters for the excitation. The absolute value of the amplitude can be obtained from the experimental results as shown in \fgref{one}b,c. The phase of the amplitude is determined with respect to the driving amplitude. As the figure clearly shows, the results are binned in two areas, where the absolute values are approximately constant, but the phases differ by $\pi$. In total we have 140 experimental runs, and 69 runs have a phase lag close to 0 and 71 runs have a phase lag close to $\pi$. This indicates that the probability $p$ to obtain a phase lag close to 0 becomes $p=0.49\pm0.04$, which is a clear signature that the symmetry breaking is indeed spontaneous with $p = 1/2$. 

In general, there can be small perturbations (noise) that can also lead to an explicit symmetry breaking. However, in our case the axial mode has a strongly oscillating spatial pattern and a well-defined oscillation frequency. This spatial and temporal pattern is impossible to induce with our magnetic coils, that are located far away from the center of our experiment. Furthermore, from the simulations that we have carried out of the process, as discussed below, we find that the axial mode grows out of an initial state  with on the order of fifty quanta reminiscent of thermal fluctuations. In the case of technical noise, the number of induced quanta can have any magnitude. Finally, for technical noise the probability $p$ can be anywhere between 0 and 1 and it is coincidental that its value becomes so close to $\infrac1/2$. 

In Ref.~\cite{Smits_2020} we have shown that our system can be very well described by the following Hamiltonian:
\begin{equation}
    \hat{H} = -\hbar \delta \had\ha + \frac{\hbar\omgD\AD}{8} ( \had\had + \ha\ha ) + \frac{\hbar g}{2} \had\had\ha\ha ,
    \label{eq:lei_4thorder}
\end{equation}%
where $\delta$ is the detuning from resonance in the rotating frame, $\omgD$ is the driving frequency, $\AD$ is the relative driving amplitude, $\ha{}^{(\dagger)}$ is the annihilation (creation) operator of a quantum in the axial mode, and $g=g'+i g''$ is a complex-valued fourth-order interaction parameter. This Hamiltonian fully describes our driven-dissipative system, where the drive is given by the term proportional to $\AD$ and the dissipation induced by the thermal cloud is determined  by the imaginary part $g''$ of the parameter $g$. The Hamiltonian in \eqref{eq:lei_4thorder} has a $\mathbb{Z}_2$ symmetry $\ha \rightarrow - \ha$, but this symmetry is spontaneously broken when $\qvalue{\ha} \neq 0$. Once the system has chosen one particular sign, the ``domino'' symmetry is broken and leads to the growth of the axial mode amplitude with this sign. 

The time-evolution of the probability distribution $P(a^*,a;t)$ of the eigenvalue of the annihilation operator is determined by a Fokker-Planck equation based on the Hamiltonian of \eqref{eq:lei_4thorder}~\cite{stoof_bijlsma2001}. Here, we simulate this numerically in a semi-classical way  solving the equations of motion for $a$ and $a^*$ (see Methods). In order to include the fluctuations due to the nonlinear dissipation, we add Stratonovich multiplicative noise with a strength given by $g''$ and a random phase, as dictated by the fluctuation-dissipation theorem. This yields a random term to the growth with a Gaussian spread. The initial distribution is $P(a^*,a;0) \propto \exp[-|a|^2/(N+1/2)]$, where $N=45$ is the number of initial thermal quanta (see Methods). For the starting value of $a(t)$ we take for each run a random initial value using this distribution. The results are similar to the experimental results, but the spread in the simulation is smaller compared to the spread in the experiment. 

Although the experimental runs are performed under identical conditions, there are always small technical fluctuations that contribute to the final result. In the experiment, superfluidity is obtained due to Bose condensation and the preparation of the superfluid droplet leads to fluctuations in the experimental parameters. There are schemes to reduce those fluctuations\cite{Kristensen_2017}, but this is beyond the scope of the present paper. Here, we can include those technical fluctuations in our simulation by taking the corresponding parameters $\delta$ and $g$ Gaussian distributed around their average value, where the spread is small compared to the average value. The results of \fgref{two} show that such technical fluctuations in combination with fluctuations due the dissipation can explain the observed width in the experiment. Note that the technical fluctuations ultimately are number fluctuations and these do not break the $\mathbb{Z}_2$ symmetry. Again the probability $p$ is close to $\infrac1/2$, showing that the SSB is well predicted by the simulation and is fully encapsulated in the model.

\begin{figure}
\includegraphics[width=0.49\textwidth]{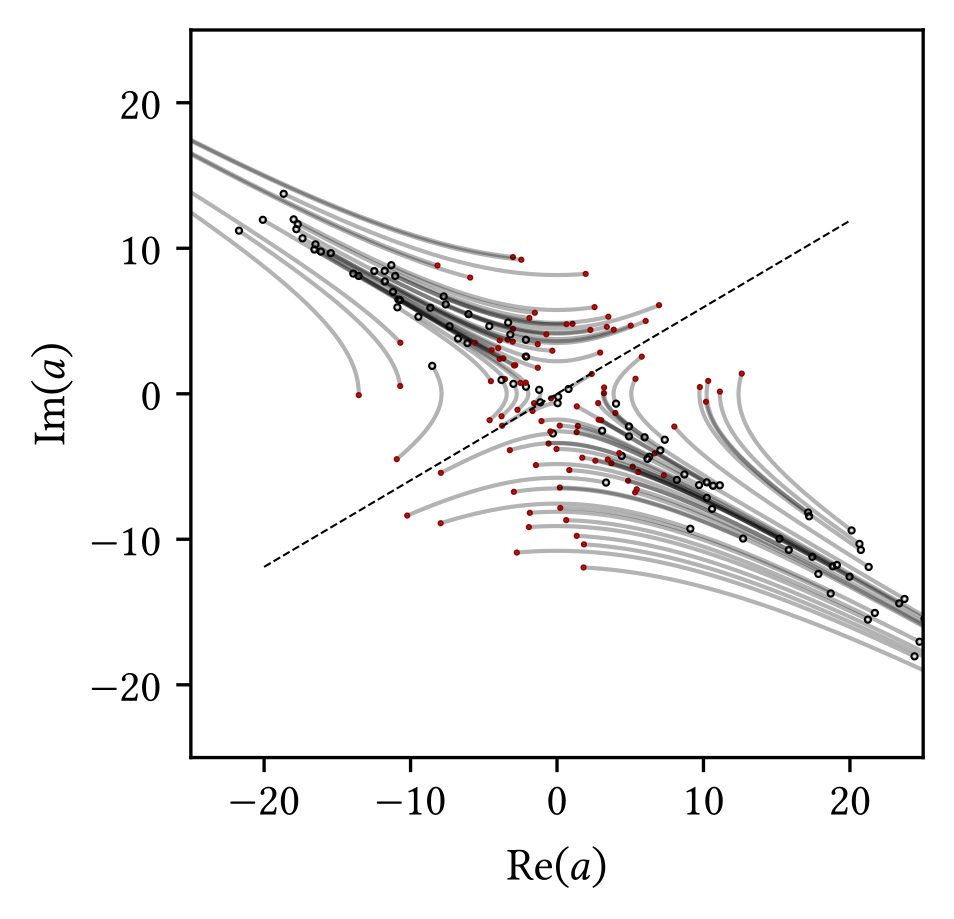}
\caption{Quantum trajectories for the expectation value $\qvalue{\ha}$, where the initial value is shown by the solid dots and the value after 10 periods by the open dots. The dashed line shows the division between initial values, where all points below the line yield a phase lag of $\phi$ and all points above the line yield a phase lag of $\phi+\pi$.  Since the initial phase in the simulation is randomly chosen, the breaking of the symmetry is spontaneous. \label{fg:four}}
\end{figure}

The simulation allows us to investigate the initial stage of the growth of $\expec{}{\ha}{}$. In \fgref{four} we show the individual trajectories for $a(t)$ for only driving the system for 10 periods. Although the initial amplitudes are fully random, all trajectories eventually evolve to either the lower right or upper left of the figure. The division is determined by the drive parameters $\delta$ and $\AD$ and the real part of the fourth-order term $g$. However, all the initial values lying below the dashed line eventually end up at a phase lag $\phi \approx 0$, whereas all initial values above this line end up at a phase lag $\phi \approx \pi$. So on the basis of a Langevin description, the breaking of the symmetry occurs due to the choice of the initial value of $a(0)$ and since this value is chosen randomly, the process is spontaneous. Of course, if the Fokker-Planck equation for $P(a^*,a;t)$ is solved directly, the symmetry is never broken explicitly. 


In conclusion, we have observed the spontaneous symmetry breaking in a driven-dissipative system. Our experiment breaks the most simple symmetry that can be broken, namely a $\mathbb{Z}_2$ symmetry, with only two possible outcomes of the experiment. Our space-time crystal is a new state of matter and allows us to further explore this symmetry breaking. For instance, by applying an excitation to the system we can induce tunneling in the system, where the phase lag will tunnel from one value to the other~\cite{PhysRevLett.126.020602}. Another possibility is to apply a $\pi/2$-type pulse to our system and drive the system from a linear superposition of both phase lags to one particular phase lag. This engineering of excitations in space and time is a rich field that requires future experiments to fully exploit all possibilities.  

\strut\\[5mm]

\section*{Acknowledgments}
We thank Dries van Oosten for valuable suggestions. The work of HS is part of the D-ITP consortium, a program of the Netherlands Organization for Scientific Research (NWO) that is funded by the Dutch Ministry of Education, Culture and Science (OCW).

\section{Methods}

\subsection{Experiment\label{sc:experiment}}
The superfluid mentioned in the main paper is a Bose-Einstein condensate of sodium atoms. Using a combination of laser cooling and evaporative cooling, a Bose-Einstein condensate of approximately $5\times10^6$ sodium atoms is created. The sodium atoms are confined in a cylindrically symmetric magnetic trap with effective trapping frequencies $(\omega_\rho,\omega_z) = 2\pi\times (92,5)\,\textrm{Hz}$. Initially the superfluid is at rest in the trap, since the evaporative cooling damps any residual motion in the superfluid due to the interaction with the thermal cloud. For the drive we only want to excite the radial breathing mode, but since all modes are coupled in the superfluid and the magnetic coils are not perfectly symmetric, many modes can become excited during the kick. The kick is induced by modulating the current through the coils providing the radial confinement. This modulation consists of two V-shaped pulses with a modulation depth of 5\% and a total duration of 10 ms per pulse. The procedure is optimized in such a way that mainly the breathing mode becomes excited. 

\begin{figure}
\includegraphics[width=0.49\textwidth]{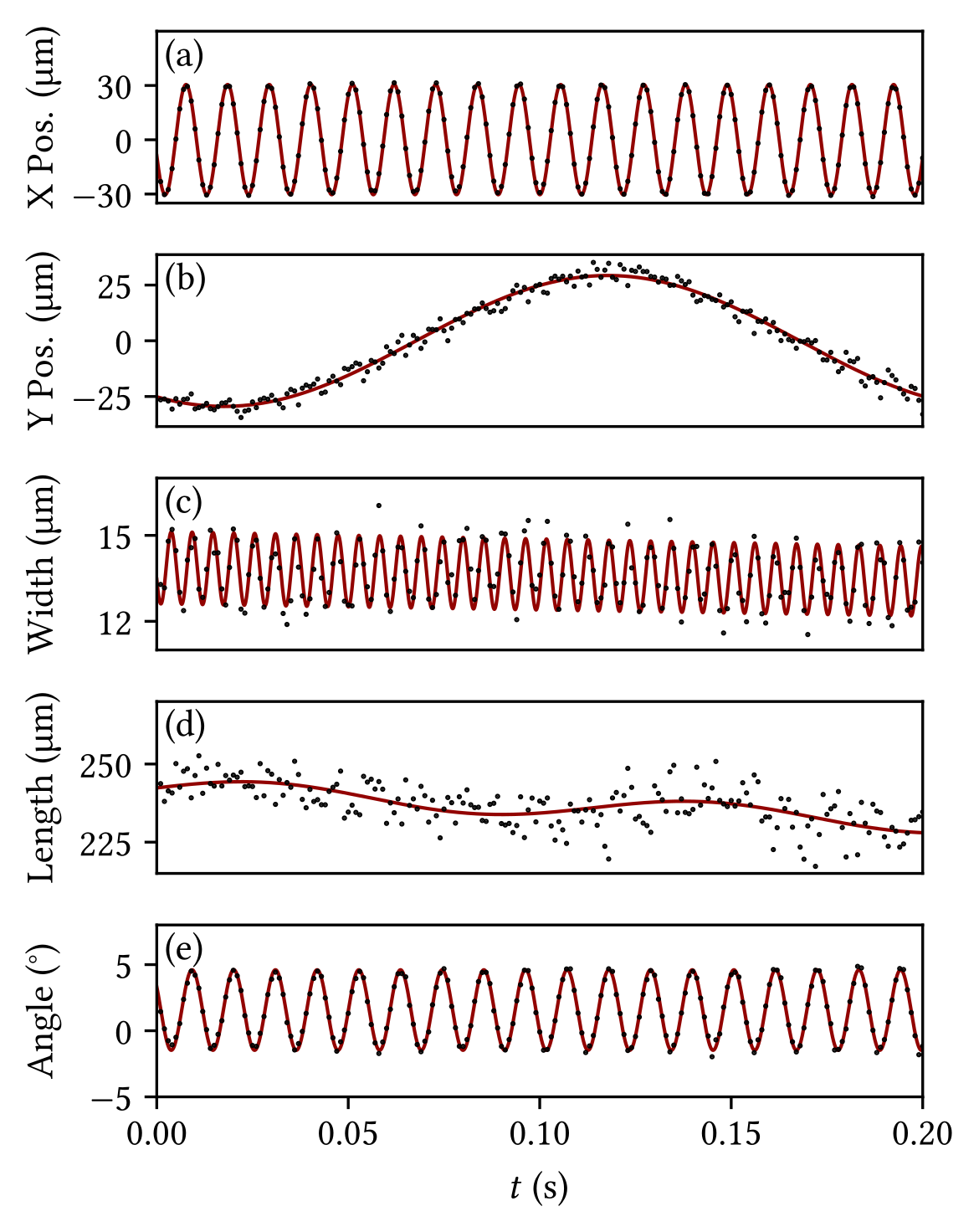}
\caption{Modes of the superfluid. {\em (a)} Position of the center of the superfluid in the radial direction. {\em (b)} Position of the center of the superfluid in the axial direction. {\em (c)} Width of the superfluid in the radial direction. (d) Width of the superfluid in the axial direction. (e) Angle of the superfluid with respect to the detection plane of the camera. \label{fg:modes}}
\end{figure}

In order to detect the different modes, the superfluid is observed using a minimally destructive holographic imaging technique. After a waiting time of $\approx$ 0.75 s over 250 images are taken, By fitting each image with a Thomas-Fermi distribution, we can determine the position, width, angle and phase shift of the superfluid. \Fgref{modes} shows the position, width and angle of the superfluid for one experimental run. In \fgref{modes}ab the position of the superfluid is shown in the radial and axial direction, respectively. The center-of-mass mode is fully decoupled from the other modes (Kohn theorem) in our harmonic trap. However, it does allow for a detection of the trap frequencies and the adjustment of the data to a sinusoidal function yields for this measurement trap frequencies of $\omega_\rho/2\pi = 92.002 \pm 0.005$ Hz and $\omega_z/2\pi = 4.98 \pm 0.03$ Hz. Note that the trap frequencies show a large ratio between the frequencies in the radial and axial direction. Also,  the oscillation of the cloud in axial direction is small.

In \fgref{modes}cd the width of the superfluid is shown. The oscillation of the breathing mode in the radial direction acts as the drive in the experiment. The frequency of the radial breathing mode is $\omega_D/2\pi = 183.74 \pm 0.09$ Hz, which is close to twice the frequency of the breathing mode as expected for a trap with a large aspect ratio. In the axial direction the breathing mode frequency is $\omega_B/2\pi = 8.1 \pm 0.2$ Hz, which is close to $\sqrt{5/2}$ of the trap frequency in the axial direction, again as expected for a trap with a large aspect ratio. Note that the amplitude of the breathing mode in the axial mode is very  small (2\% of the width). If the axial width oscillates strongly in time, the resonance condition for the high-order mode in the axial direction depends on time and thus not one mode becomes excited. In our case, the excitation scheme causes only a single mode to become excited.

In \fgref{modes}e the angle of the superfluid is shown. The angle oscillates in time and this is due to the scissor mode in the superfluid~\cite{scissor}. The scissor mode is a clear sign of the superfluidity in the fluid and the frequency is $\omega_{\rm sc}/2\pi = 91.887 \pm  0.014$ Hz. The frequency of the scissor mode is close to the trap frequency in the radial direction, since the axial frequency is much smaller than the radial frequency. Note that the amplitude of the scissor mode is constant and that the spatial symmetry of the scissor mode is uneven with respect to reflection in the plane containing the long axis of the condensate, and thus does not couple to the high-order axial mode.

\begin{figure}
\includegraphics[width=0.49\textwidth]{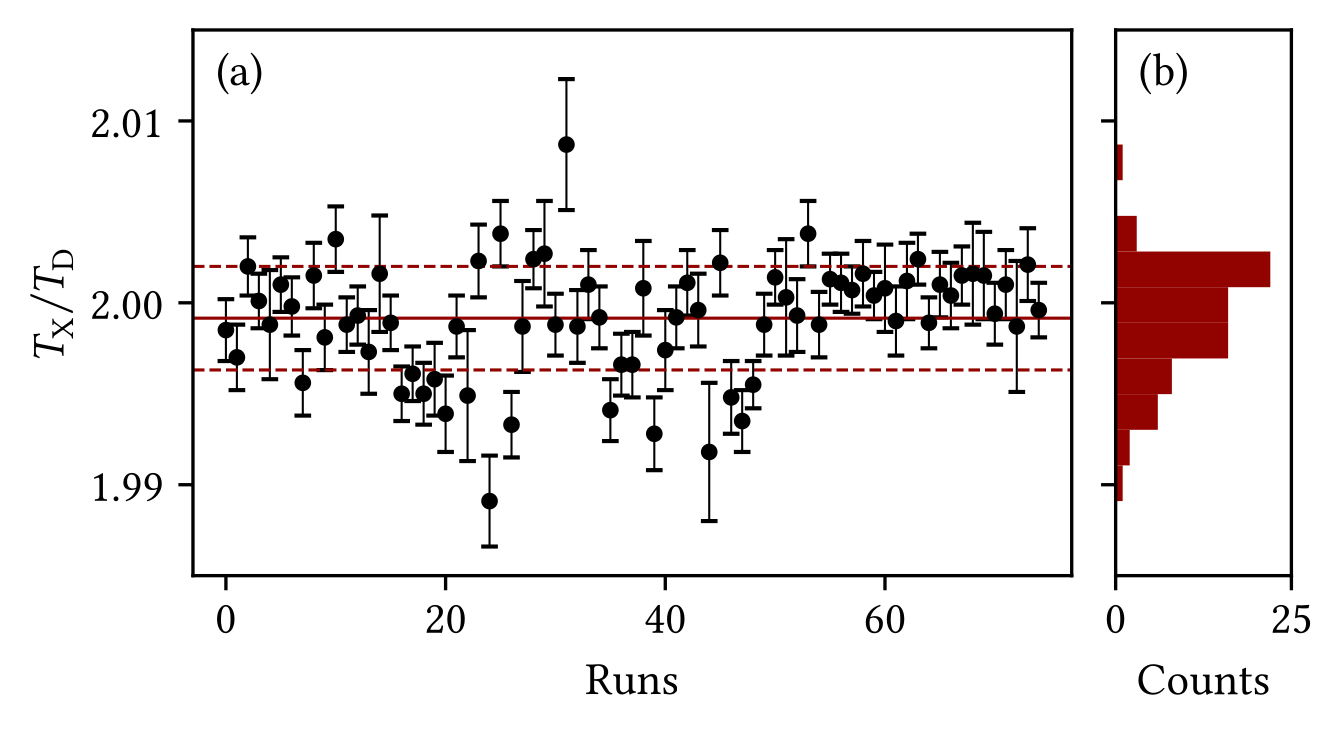}
\caption{The ratio between the period $T_X$ of the high-order axial mode and the period $T_D$ of the drive for all the runs on a day. The ratio is 1.999 $\pm$ 0.003, which is within 1\textperthousand\ of the factor 2 to be expected for a discrete time-crystal.   \label{fg:dtc}}
\end{figure}

In \fgref{dtc} the ratio between the period of the high-order axial mode and the period of the drive is shown for the experimental runs on one day. The figure shows that the ratio is exactly two within experimental uncertainty, as expected for the discrete time crystal~\cite{10.1146,sacha-review}. The results show that the SSB experiment can be carried out in a reproducible way over a long period of time. 

\begin{figure}
\includegraphics[width=0.49\textwidth]{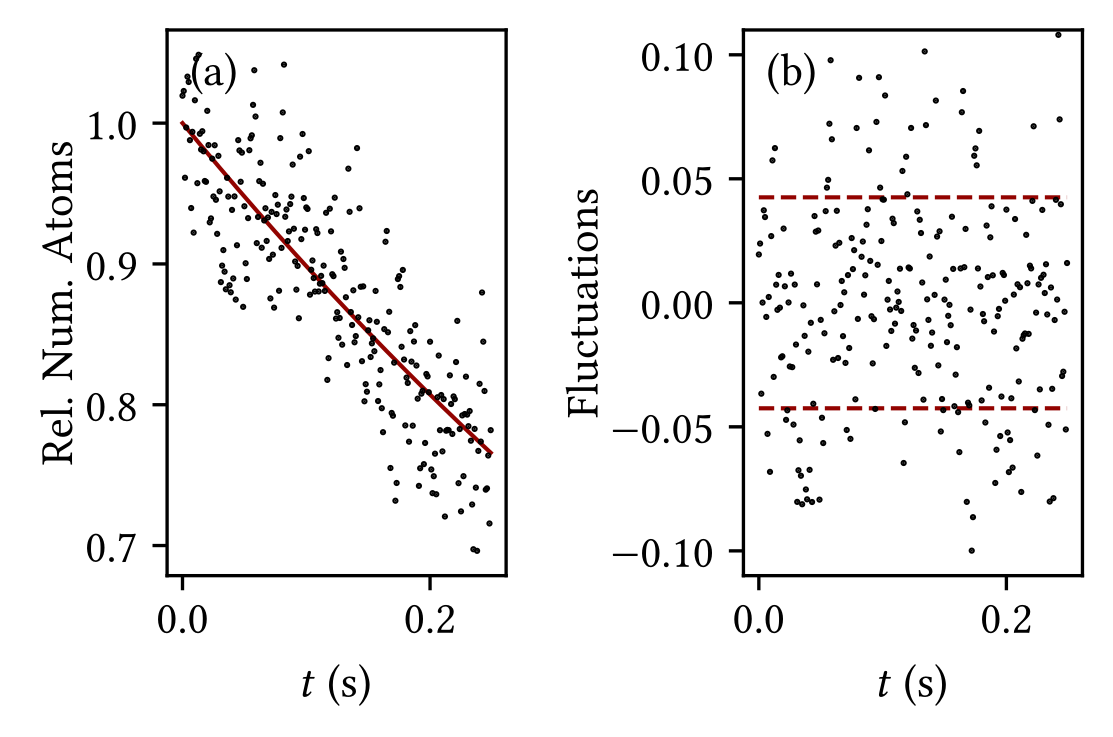}
\caption{Number for atoms in the superfluid. {\em (a)} Total number of atoms in the superfluid as a function of time during the imaging. The number of atoms exponentially decreases with a time constant determined by the imaging technique. {\em (b)} Fluctuation of the number of atoms as a function of time.  \label{fg:atoms}}
\end{figure}

In the off-axis holography we also detect the phase shift of the center of the superfluid, and together with the width of the superfluid in the axial and radial direction we can determine the total number of atoms in the superfluid. In \fgref{atoms}a the number of atoms is shown as function of the time during the detection. The time constant $\tau$ for decay due to the imaging is $\tau=0.93 \pm 0.04$ s, and since the dwell time between images is 1 ms, the loss rate per images is only 0.09\%, which is extremely low. In \fgref{atoms}b the fluctuations in the imaging between subsequent images is shown, corrected for the exponential decay caused by the imaging. The results show that the statistical uncertainty in the detection technique is 4\%, which makes the technique very reliable for imaging the superfluid.
 
\subsection{Analysis}
From each measurement run, the experimental data is analyzed as described in previous work~\cite{Smits_2020}. The images are fitted with a function which contains the density profile of the high-order axial mode. From the fit we obtain the position and widths of the cloud, and the amplitude $A_X$ of the axial profile, for which the results are shown in \fgref{one}. From the fit of the breathing mode, we choose a zero crossing of the oscillation with positive ramp at a time $t_0$ after a fixed number of oscillations after the kick. We identify the zero crossing in the fit of the axial mode closest to $t_0$ and determine the phase lag $\phi$ based on the time between the zero crossing and $t_0$. In case of a zero crossing with negative ramp, we add an additional factor $\pi$ to the phase lag. This method properly takes into account the difference between $\omega_D$ and $2 \omega_X$ in the fit. 

\begin{figure}
\includegraphics[width=0.49\textwidth]{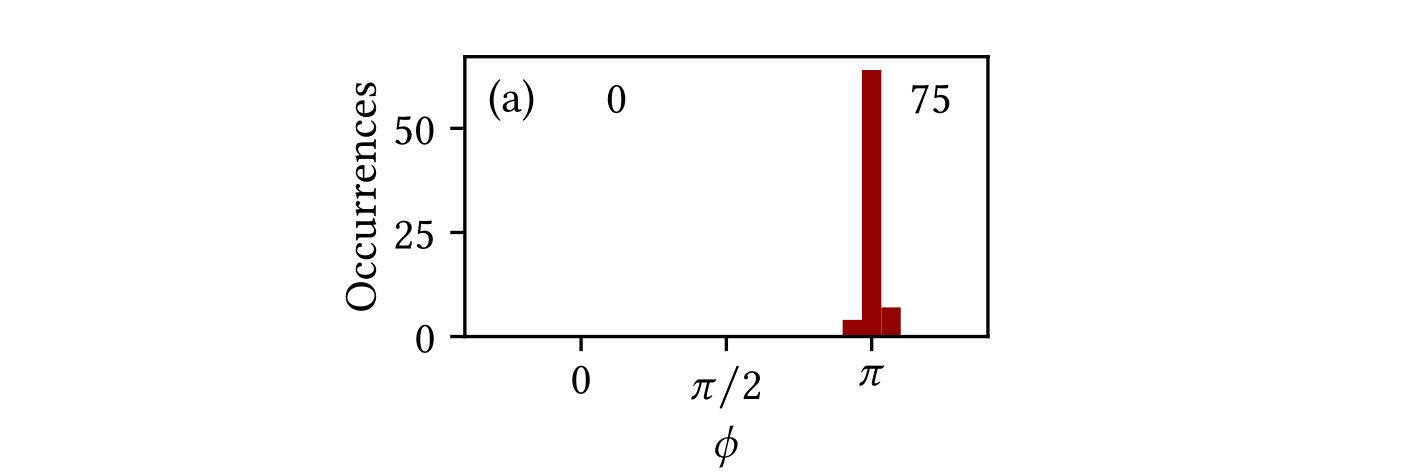}
\caption{Histogram of the phase lag between the scissor mode and the drive. \label{fg:scissor}}
\end{figure}

To rule out the possibility that the high-order axial mode is driven by the scissor mode, we have determined the phase lag of the scissor mode with respect to the drive using the same method. The scissor mode is found to be always pinned at approximately a phase lag of $\pi$, as shown in \fgref{scissor}. This precludes the possibility that the scissor mode drives the high-order axial mode.

\subsection{Number of quanta\label{sc:quanta}}
The high-order axial mode is quantized, as shown in Ref.~\cite{liao19}. The density of the mode $n_X(z,t)$ in the axial direction is given by $n_X(z,t) \equiv - \dot{\kappa}(t) L_j(\bar{z})$, where the dot represents the time-derivative and $L_j(\bar{z}) =P_j(\bar{z})-P_{j-2}(\bar{z})$ is the mode profile and $j$ the mode number. Here $\bar{z}=z/R_z$ is the reduced distance in the $z$-direction with $R_z$ the Thomas-Fermi width and $P_j(\bar{z})$ are Legendre polynomials of order $j$. It is assumed that there is no dependence of the mode in the radial direction and that the density of the axial profile is only non-zero, where the superfluid has a non-zero density.  In the frame rotating with the drive frequency $\omega_D$, the amplitude $\tilde{\kappa}$ can be related to the annihilation operator $\ha$ using $\tilde{\kappa} = q \ha$, where $q=\sqrt{{\hbar}/({\eta Q \omega_D})}$ is the normalization parameter. Here $Q$ is the overlap of the mode function, 
$$ Q = \int_{-1}^{+1} \dd \bar{z} (1-\bar{z}^2) L_j(\bar{z})^2, $$
and $\eta=\pi T^{\rm 2B} R_\rho R_z$ is the effective mass parameter. Using the experimental parameters the density modulation for one quantum in the axial mode becomes {1.03}$\times10^{15}$ atoms/m$^3$, which is far too small to be observable in our setup. However, the axial mode grows exponentially in time and after a wait time of about 1 s, the density modulation has become sufficiently large to be observable. This analysis allows us to scale the value of $a$ from the simulation to the amplitude $A_X$ and plot its value in \fgref{two}. The density modulation is proportional to the time-derivative of $\kappa$ and thus we have to add a factor $\pi/4$ to the phase of $\kappa$ to compare our results with the experimental values.

\subsection{Initial number of quanta}
The initial number of quanta $N$ in the axial profile is determined by the thermal fraction of this mode, as determined by the Bose-Einstein distribution. The temperature of the thermal cloud is in our case $T \approx 200$  nK and the energy of the mode becomes $\hbar \omega_X$, where $\omega_X/2\pi$ = 92 Hz is the frequency of the high-order axial mode. So the number of quanta becomes $N_{\rm init} = 45$, which is much larger than the quantum fluctuation of \infrac1/2.

\subsection{Fluctuations in the growth}
The semi-classical analysis starts with the equations of motion for $a(t)$ and $a(t)^*$ as given in Ref.~\cite{Smits_2020}:
\begin{equation}
i \frac{\textrm{d}}{\textrm{d}t} a = \left( -\delta + g|a|^2 \right) a + \frac{\omgD\AD}{4} a^* \label{eq:a_eom}, 
\end{equation}
and the complex conjugate for $a^*(t)$. In order to include the fluctuations we have to include multiplicative noise $\eta(t) a^*(t)$ with $\langle \eta^*(t) \eta(t) \rangle = D \delta(t-t')$ in the  model, where $D$ in our case is given by $D = 2 \hbar (N_{\rm fluc}+\infrac1/2) g''$ as determined by the fluctuation-dissipation theorem. Here, $N_{\rm fluc}$ accounts for the increase of noise due to thermal fluctuations. Its value is estimated by first determining the oscillation frequency $\omega_{\rm eff}$ in the effective potential $V(|a|)$, as given in Eq.~(7) of Ref.~\cite{Smits_2020}. Given our experimental parameters we find $\omega_{\rm eff}$ = 8.5 Hz and assuming that our system is in equilibrium with the thermal cloud at a temperate $T$ = 200 nK, the number of quanta becomes $N_{\rm fluc}$ = 456. 

To include the fluctuations to the model we add a stochastic noise term $a^*(t) \sqrt{D/\Delta t} \, x_i$ to \eqref{eq:a_eom} and the complex conjugate to the equation for $a^*(t)$ with $\Delta t$ the step-size in time in the integration. Here, $x_i$ is a Gaussian-distributed complex variable with unit absolute value, which is randomly selected for each time step.

\begin{figure}
\includegraphics[width=0.49\textwidth]{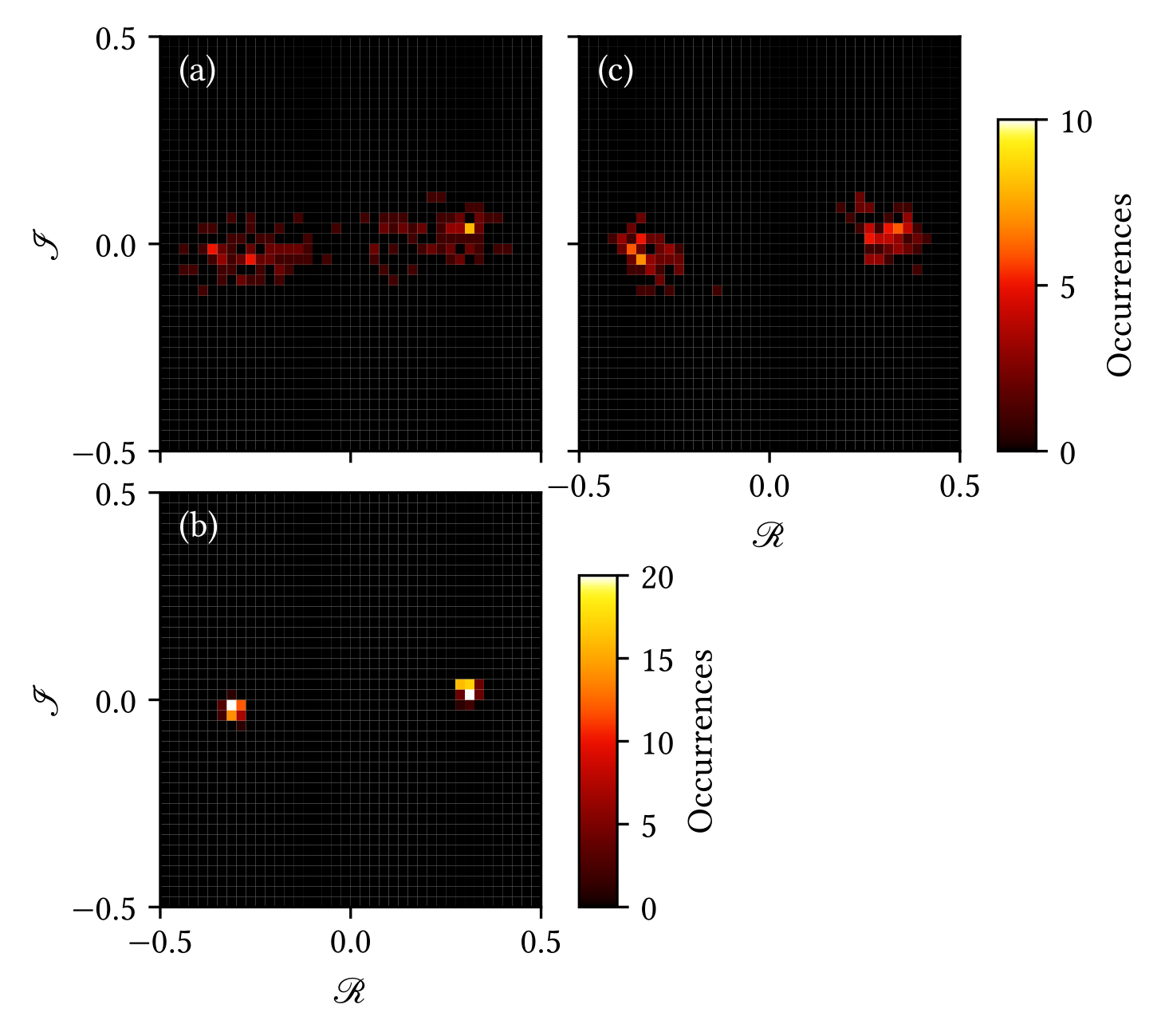}
\caption{Comparison between the experiment and the simulations for the high-order axial mode amplitude $A_X$. In each case the histogram contains 140 runs. {\em (a)} Contour plot of the results of the experiments. {\em (b)} Contour plot of the simulations, where only the thermal fluctuations are taken into account. {\em (c)} Contour plot of the simulations, where both the thermal and technical fluctuations are taken into account.   
\label{fg:2d_histo}}
\end{figure}

\subsection{Dependencies on system parameters}

\begin{figure}
\includegraphics[width=0.49\textwidth]{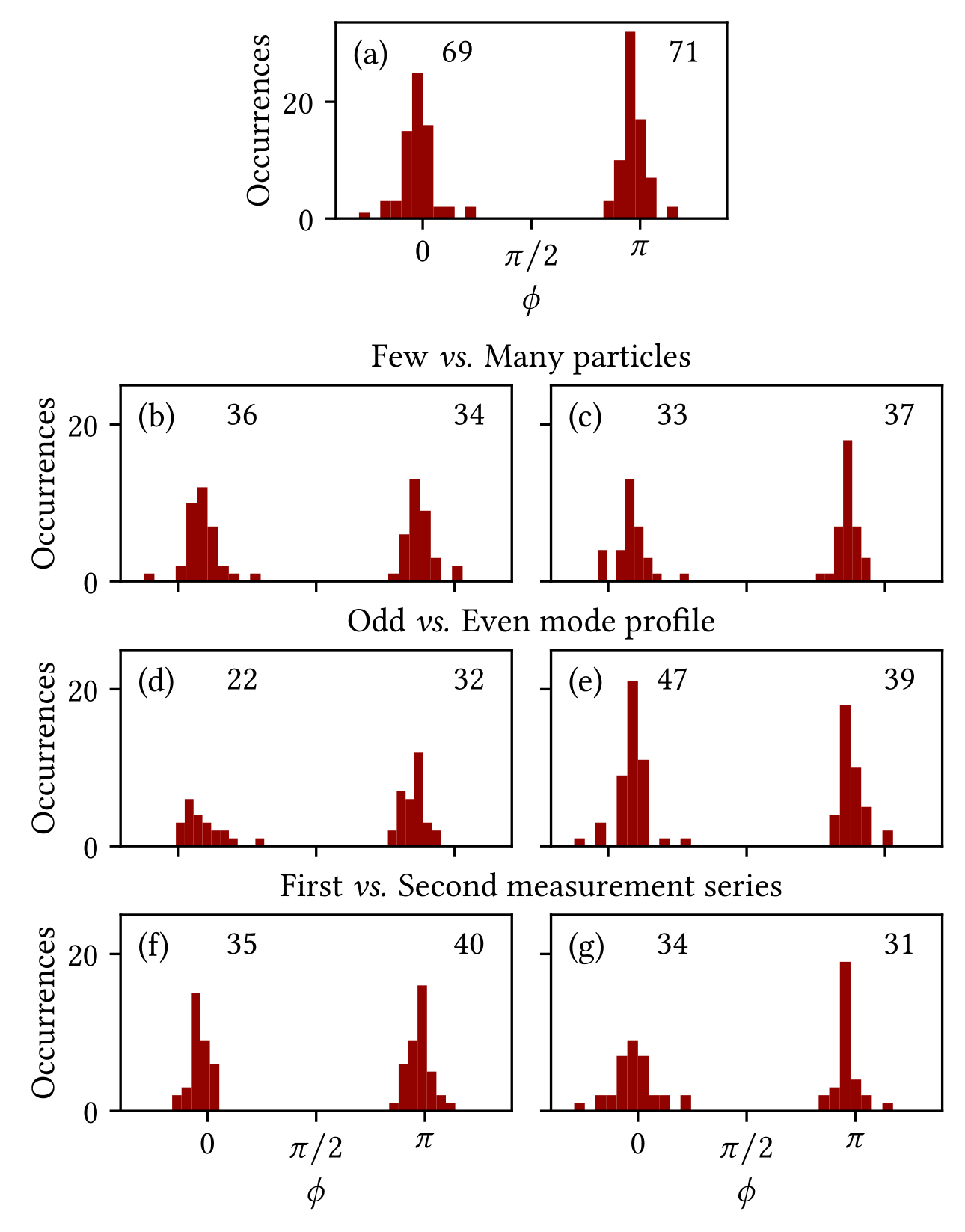}
\caption{Histogram of the phase with data split in different ways. By splitting the data set it is excluded that $\phi$ or $\phi+\pi$ is more prevalent in certain subsets of the data set. It is seen that, cutting the data set into approximately equal pieces does not disturb the 50/50 balance. \emph{(a)} Full data set. \emph{(b,c)} Data set split into less \emph{(b)} and more \emph{(c)} particles than the median number of particles in the BEC. \emph{(d,e)} Data split into odd \emph{(d)} and even \emph{(e)} modes, as described in previous work~\cite{Smits_2020}. \emph{(f,g)} First \emph{(f)} and second \emph{(g)} measurement series. \label{fg:barplotsplit}}
\end{figure}

To exclude the possibility of the phase being pinned at either $\phi$ or $\phi+\pi$ by some underlying phenomena related to particle number, mode function or a date-specific parameter, the data set has been split in two parts along different criteria, see \fgref{barplotsplit}. \fgref{barplotsplit}a shows the same histogram as in the main paper. The chance to get phase $\phi$ is calculated from the data as $p = {N_0}/({N_0+N_\pi})$, with uncertainty $\sigma_p = \sqrt{{p(1-p)}/({N_0+N_\pi})}$. For the full data set, this results in $p = 0.49 \pm 0.04$. \Fgref{barplotsplit}bc show the data split by particle number. In \fgref{barplotsplit}b, all data points with a particle number below the median particle number are taken, which results in $p = 0.51\pm0.06$. In \fgref{barplotsplit}c, all data points with a particle number above the median particle number are taken, which results in $p = 0.47\pm0.06$. As calculated probabilities are within a margin of error of $p=1/2$, the particle number does not appear to prefer one solution for the phase over the other. \Fgref{barplotsplit}de show the data split by mode number (see \scref{quanta}). In \fgref{barplotsplit}d, all data points with an odd mode are shown, which results in $p = 0.41\pm0.07$. In \fgref{barplotsplit}e, all data points with an even mode are shown, which results in $p = 0.55\pm0.05$. For odd modes, the calculated probability deviates from $p$ = \infrac1/2 with a  margin of error which is a little large than one standard deviation, however, the number of data points is small. Finally, since data was acquired on two separate days, the data set was split in \fgref{barplotsplit}fg by measurement series. Data acquired on the first day is shown in  \fgref{barplotsplit}f. Analysis of this data results in $p = 0.47\pm0.06$. Data acquired on the second day is shown in  \fgref{barplotsplit}g. Analysis of this data results in $p = 0.52\pm0.06$. From this it is concluded that from day to day there no preference between either the $\phi$ or $\phi+\pi$ solutions. 

{\it Note added after completion of this work:} In the appendix of Ref.~\cite{Kessler} we have found in the Methods section also experimental results for the breaking of a different $\mathbb{Z}_2$ symmetry in a driven-dissipative system.

\bibliographystyle{apsrev}
\bibliography{SSBNotes}

\end{document}